\documentclass[12pt]{article}
\usepackage{amsmath,latexsym,epsfig,amssymb}

\input eqalign.sty     

\newcommand{\dds}{\stackrel{\leftrightarrow}{D}}

\newcommand\be{\begin{equation}}
\newcommand\ee{\end{equation}}
\newcommand\bea{\begin{eqnarray}}
\newcommand\eea{\end{eqnarray}}

\def\ma[#1,#2,#3,#4]  {{\left( \matrix{ #1  & #2 \cr
                                        #3  & #4 \cr } \right)}}

\begin{document}

\title{
{\vspace{-1cm} \normalsize
\hfill \parbox{40mm}{CERN/TH-99-10}}\\
Universal continuum limit of non-perturbative lattice non-singlet moment evolution }
\author{
Marco Guagnelli$^1$,  
Karl Jansen$^{2,}$\footnote{Heisenberg Foundation Fellow}  
$\;$ and Roberto Petronzio$^{1,}$$^2$  
\\
{\footnotesize $^1$ Dipartimento di Fisica, Universit\`a di Roma 
{\em Tor Vergata} }\\
{\footnotesize and INFN, Sezione di Roma II} \\
{\footnotesize Via della Ricerca Scientifica 1, 00133 Rome, Italy } \\
{\footnotesize 
$^2$ CERN, 1211 Geneva 23, Switzerland  
}
}
\maketitle

\begin{abstract}
We present evidence for the universality of the
continuum limit of the scale dependence of the renormalization
constant
associated with the operator
corresponding to the average momentum of non-singlet parton
densities. 
The evidence is provided by a non-perturbative computation
in quenched lattice QCD 
using the Schr\"odinger Functional
scheme. 
In particular, we show that the continuum limit is independent of
the form of the fermion action used, i.e. the Wilson action and the
non-perturbatively improved clover action.
\end{abstract}

\pagebreak

We presented in ref.~\cite{our_non_pert} a 
quenched non-perturbative calculation  
for the renormalization constant associated with 
the operator
corresponding to the average momentum of non-singlet parton densities.
The calculation was         
based on
a finite size recursive scheme that allows us to reconstruct in the
continuum the non-perturbative evolution of the renormalization
constant in the Schr\"odinger Functional (SF)
scheme~\cite{schrfunc,sint,paper3,letter}. 
In  ref.~\cite{our_non_pert} we used a standard Wilson action where the
continuum limit is approached with a rate that is linear in the lattice
spacing $a$ in leading order. The results for the step
scaling functions presented in ref.~\cite{our_non_pert} also showed
sizeable quadratic corrections that had to be included in the fit and
made the continuum extrapolation more hazardous. 

We repeated the same
calculation with a non-perturbatively improved clover
action~\cite{paper3} that
changes the form of the lattice artefacts. Linear corrections are
still to be expected, because the ${\rm O}(a)$-improvement of the action should be
complemented with the improvement of the operators and of the boundary
counterterms in order to lead to a full cancellation of effects appearing linear
in $a$. However, the continuum limit cannot depend upon the
discretization chosen and should be universal. Our results do indeed
show such a universality and 
therefore put our previous
continuum extrapolations on a firmer basis.
We find that the extrapolation 
of the step scaling function, computed with the ${\rm O}(a)$-improved
action, to the contiuum need
again quadratic terms in the fit.                    

We refer to ref.~\cite{our_non_pert} for more of the details about the
calculation which we here only shortly summarize as follows.  
We calculate the
renormalization constant of the twist-two non-singlet operator defined
by:
\begin{equation}
{\cal{O}}_{\mu_{1}\ldots \mu_{n}}^{qNS} = 
\bigl( \frac{i}{2}\bigl)^{n-1} {\bar{\psi}}(x)\gamma_{\{\mu_{1}} 
\dds_{\mu_{2}}\cdots \dds_{\mu_{n}\}} 
\frac{\lambda^f}{2} \psi(x)\ +\ \mbox{trace terms}\; .
\label{eq:twist_two_continuum}
\end{equation}
The basic ingredient for the reconstruction of the non-perturbative
scale dependence of the renormalization constants of the 
above operator is the finite-size step scaling function $\sigma_Z$ defined by:
\begin{equation}\label{Zs}
Z(sL) = \sigma_Z(\bar{g}^2(L))Z(L)\; ,
\end{equation}
where $L$ is the physical length that plays the role of the
renormalization scale and $Z$ the renormalization constant of the
operator, which is
defined by:
\begin{equation}
O^{R}(\mu) = Z(1/a\mu)^{-1}O^{\rm bare}(a/L)\; .
\end{equation}
$Z$ is
obtained from the SF matrix element of the operator on a finite volume
$L^3T$ normalized to its tree level:
\begin{equation}
O^{\rm bare}(a/L) = Z(L/a) O^{\rm tree}\; .
\end{equation}
The framework of the Schr\"odinger Functional, which describes the
quantum time evolution between two fixed classical gauge and fermion
configurations defined at time $t=0$ and $t=T$, has been used
extensively in the recent literature 
\cite{paper4,qmass} to calculate non-perturbative
renormalization constants of local operators.  Among the advantages of
the method, we only quote the possibility of performing the
computations at zero physical
quark mass and of using non-local gauge invariant sources for the fermions 
without need of a gauge-fixing procedure. In our particular case, we
exploit both features. Our observable is defined by~\cite{roberto_pert}:
\begin{equation}
Z =
\frac{f_2(x_0=L/4)}{\sqrt{f_1}}\left/\left(\frac{f_2(x_0=L/4)}{\sqrt{f_1}}\right)_{\rm
tree}\right.\; ,
\label{eq:obs}
\end{equation}
with $f_2$ given by
\begin{equation}
f_2(x_0) = -a^6\sum_{\bf{y},\bf{z}} \rm{e}^
{i\bf{p}(\bf{y}-\bf{z})}
\langle \frac{1}{4} \bar\psi(x) \gamma_{\{1} 
\dds_{2\}}\frac{1}{2} \tau^3 \psi(x) 
\bar\zeta({\bf{y}}) \Gamma \frac{1}{2} \tau^3 \zeta({\bf{z}})\rangle
\label{eq:f2}
\end{equation}
and $f_1$ by
\begin{equation}
f_1 = -a^{12}\sum_{\bf{y},\bf{z},\bf{v},\bf{w}}\langle
\bar\zeta'(\bf{v})\frac{\tau^3}{2}\zeta'(\bf{w})
\bar\zeta(\bf{x})\frac{\tau^3}{2}\zeta(\bf{y})\rangle
\label{eq:f1}
\end{equation}
where $\zeta=\delta /\delta\bar{\psi}_c$ and $\bar\zeta=-\delta/\delta\psi_c$ 
are the derivatives with respect to the
two-component classical fermion fields ($\bar{\psi}_c$ and $\psi_c$,
respectively) at the boundary $x_0=0$, while $\zeta'$ and $\bar\zeta'$
are the corresponding derivatives at the boundary $x_0=T$. The
projection on the classical components is achieved by the projector
$P_{\pm}$ defined by $\frac{1}{2}(1\pm\gamma_0)$. On the boundaries the theory possesses 
only a
{\em global} gauge invariance that is preserved by the
quantities defined above.  The values of $x_0$ (set to $T/4$) and of the 
non-zero component of the momentum $p_x$ (set to $2\pi$/$L$) are both
scaled in units of $L$, which therefore remains the only scale 
besides the lattice spacing $a$. The quantity $f_1$
serves as a normalization factor that removes the wave function
renormalization constant of the $\zeta$ fields in order to isolate the
running associated with the operator in
eq.~(\ref{eq:twist_two_continuum}) only.

In order to compute the running of $Z$ in eq.~(\ref{eq:obs}) its 
step scaling function $\sigma_Z\equiv \sigma_{\bar{Z}}/\sigma_{f_1}$
with 
\begin{equation}
\label{eq:sigmas}
\sigma_{\bar{Z}}= \frac{f_2(x_0=L/4)}{\left(f_2(x_0=L/4)\right)_{\rm tree}}
\; ;\;\; \sigma_{f_1} =\frac{\sqrt{f_1}}{\left(\sqrt{f_1}\right)_{\rm tree}}
\end{equation}
has to be evaluated in the continuum. 
To this end, 
the physical size $L$ ($= T$ in our case) is kept fixed in
physical units by keeping fixed the renormalized coupling constant in
the SF scheme, renormalized at the physical volume scale.  By
increasing the number of lattice points while keeping the value of $L$
constant, the lattice spacing in physical units is reduced and a
continuum extrapolation of the step scaling function can be performed.

The presence of a large non-zero momentum induces potentially large
lattice artefacts in the numerator of the renormalization constant in
eq.~(\ref{eq:obs}). 
Indeed, in ref.~\cite{our_non_pert}, for this reason, we had to extrapolate
independently the step scaling functions $\sigma_{\bar{Z}}$ and for $\sigma_{f_1}$, with a
quadratic and a linear fit in the lattice spacing, respectively.

\begin{figure}
\vspace{0.0cm}
\begin{center}
\psfig{file=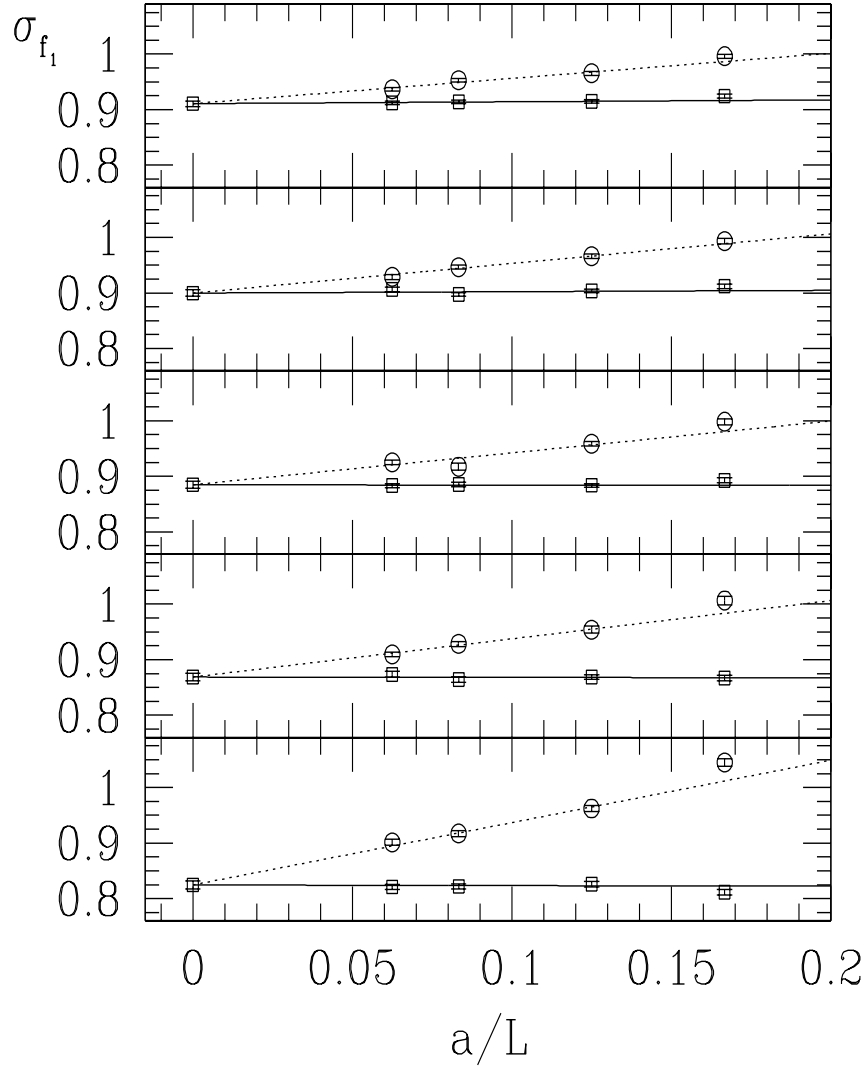, %
width=13cm,height=16cm}
\end{center}
\caption{ \label{fig:f1comb} Continuum extrapolation of
$\sigma_{f_1}$ using a combined fit at all values of $\bar{g}^2$ of table~1, with 
$\bar{g}^2$ increasing from top to bottom. 
Circles and the dashed line correspond to using Wilson fermions.
Squares and the solid line correspond to using the
${\rm O}(a)$-improved action.
The fit for $\sigma_{f_1}$ is linear in both cases using the three points with smallest
$a/L$. 
}
\end{figure}

\begin{figure}
\vspace{0.0cm}
\begin{center}
\psfig{file=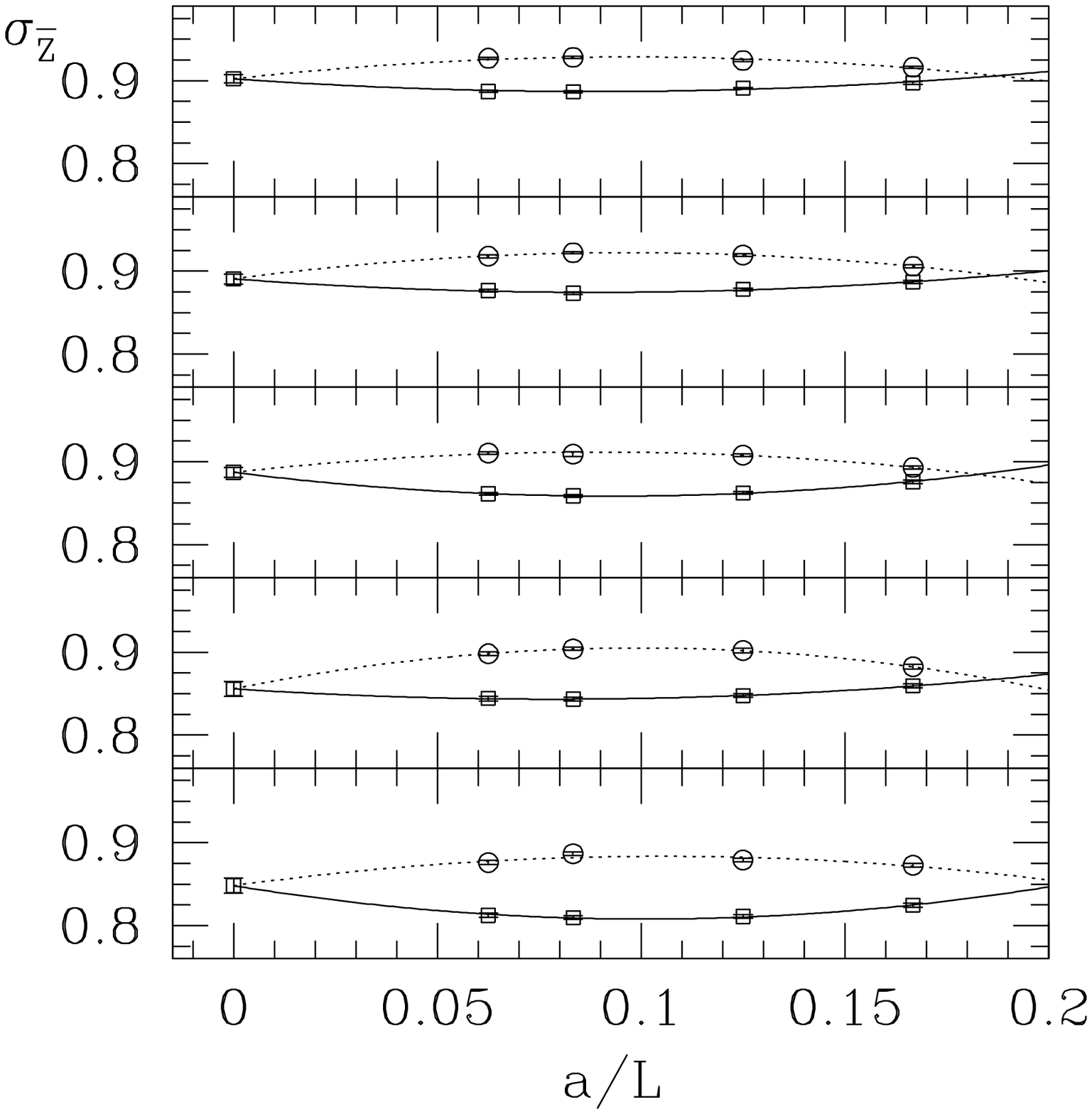, %
width=13cm,height=16cm}
\end{center}
\caption{ \label{fig:zbarcomb} 
Continuum extrapolation of
$\sigma_{\bar{Z}}$ using a combined fit at all values of $\bar{g}^2$ of table~2, with 
$\bar{g}^2$ increasing from top to bottom. 
The fit for $\sigma_{\bar{Z}}$ is quadratic using
all data points.
Notation is as in fig.~1.
}
\end{figure}

As already mentioned, the calculation done with a non-perturbatively
improved action, without improving the boundary terms and the
operator (whose matrix element in our case is tree-level-improved) does not
eliminate all linear lattice artefacts and certainly does not
eliminate the quadratic artefacts, if present with the Wilson
action. Therefore, also with the clover action we have done
independent extrapolations for $\sigma_{\bar{Z}}$ and for $\sigma_{f_1}$, 
which are
compared with our previous results using the 
Wilson action in tables 1 and 2 for the different
values of $L$ (or equivalently $\bar{g}^2(L)$) considered in ref.~\cite{our_non_pert}.
Within the errors, there is full consistency between the two sets of
results obtained with different fermion actions, showing the
universality of the continuum limit. 
Given the consistency of the extrapolated values of the step scaling
functions from both fermion actions, we performed a combined fit,
using both data sets obtained with the two fermion actions. 
In the combined fit we demanded that the extrapolations go to the
same continuum value of the step scaling function. 
In figures \ref{fig:f1comb} 
and \ref{fig:zbarcomb} we report
results at fixed non-zero lattice spacing used for the extrapolation
and the corresponding combined fit: while the linear dependence upon
the lattice artefacts for $\sigma_{f_1}$ is almost eliminated, the quadratic
terms for $\sigma_{\bar{Z}}$ are still needed, as expected. 
In tables
\ref{table_fit_f1} and \ref{table_fit_zbar} we report
the $\chi^2$ per degree of freedom (d.o.f.) values of the combined fit with those of the original fit
of ref.~\cite{our_non_pert}
and a fit for the case of the improved action alone.
The $\chi^2$ values are totally comparable, as are our final values for the continuum
step scaling function. Note that the error compared with the ones given in ref.~\cite{our_non_pert}
gets decreased by about a factor $\sqrt 2$.

\vspace{0.3cm}

We have shown that the continuum extrapolation of the evolution of the
renormalization constant corresponding to the average momentum of
non-singlet parton densities shows the expected independence upon the
lattice action used. Over the range of scales explored, the evolution
is adequately described with a three loop expression for the anomalous
dimensions \cite{our_non_pert}. Further simulations that extend the running into the
perturbative region, where a good description in terms of the known
perturbative evolution can be made, are under way.

\pagebreak


\begin{table}[htbp]
\begin{center}
\leavevmode
\begin{tabular}[]{|c|c|c|c|c|c|c|}
\hline
$\bar{g}^2(L)$&$\sigma_{f_1}({\rm Wilson})$&$\chi^2$&
               $\sigma_{f_1}({\rm impr})$&$\chi^2$&
               $\sigma_{f_1}({\rm combined})$&$\chi^2$ \\
\hline\hline
 1.8811       & 0.912(7)  & 2.61  & 0.910(7) & 0.06  & 0.911(5) & 0.90 \\
 2.1000       & 0.897(9)  & 1.17  & 0.902(8) & 3.22  & 0.900(6) & 1.54 \\
 2.4484       & 0.886(9)  & 6.90  & 0.884(7) & 0.17  & 0.885(6) & 2.36 \\
 2.7700       & 0.866(10) & 0.52  & 0.871(10)& 1.58  & 0.869(7) & 0.73 \\
 3.48         & 0.837(11) & 0.39  & 0.815(10)& 0.01  & 0.825(7) & 0.86 \\
\hline    
\end{tabular}
\caption{
         The values of the step scaling function 
         $\sigma_{f_1}$ are given extrapolated to the continuum.  We
         compare the cases for Wilson fermions (2nd column), using the
         ${\rm O}(a)$-improved action (4th column) and from the
         combined fit (6th column).  The running coupling is computed
         in the SF scheme.  We give also the $\chi^2/{\rm d.o.f}$.}
\label{table_fit_f1} 
\end{center}
\end{table}


\begin{table}[htbp]
\begin{center}
\leavevmode
\begin{tabular}[]{|c|c|c|c|c|c|c|}
\hline
$\bar{g}^2(L)$&$\sigma_{\bar{Z}}({\rm Wilson})$&$\chi^2$&
               $\sigma_{\bar{Z}}({\rm impr})$&$\chi^2$&
               $\sigma_{\bar{Z}}({\rm combined})$&$\chi^2$ \\
\hline\hline
 1.8811       & 0.917(8) &0.19 & 0.890(7) & 0.20  & 0.902(5) &2.55 \\
 2.1000       & 0.891(9) &0.20 & 0.891(9) & 0.89  & 0.891(6) &0.36  \\
 2.4484       & 0.891(9) &1.01 & 0.884(9) & 0.03  & 0.887(6) &0.43 \\
 2.7700       & 0.854(13)&0.01 & 0.858(13)& 0.002 & 0.856(9) &0.02 \\
 3.48         & 0.855(12)&6.44 & 0.840(13)& 0.04  & 0.848(9) &2.42 \\
\hline
\end{tabular}
\caption{
         The values of the step scaling function   
         $\sigma_{\bar{Z}}$ are given extrapolated to the continuum.
         We compare the cases for Wilson fermions (2nd column), using
         the ${\rm O}(a)$-improved action (4th column) and from the
         combined fit (6th column).}
\label{table_fit_zbar}
\end{center}
\end{table}

\pagebreak

\begin{thebibliography}{99}
%
\bibitem{our_non_pert}
M.~Guagnelli, K.~Jansen and R.~Petronzio, 
hep--lat/9809009, to be published in Nucl. Phys. B.
%
\bibitem{schrfunc} 
M.~L\"uscher, R.~Narayanan, P.~Weisz and U.~Wolff,
Nucl. Phys. B384 (1992) 168.
%
\bibitem{sint}
S.~Sint, 
Nucl. Phys. B421 (1994) 135.
%
\bibitem{paper3} 
M.~L\"{u}scher, S.~Sint, R.~Sommer, P.~Weisz and U.~Wolff,
Nucl. Phys. B491 (1997) 232.
%
\bibitem{letter} 
K.~Jansen, C.~Liu, M.~L\"{u}scher, H.~Simma, S.~Sint, R.~Sommer,
P.~Weisz and U.~Wolff, 
Phys. Lett. B372 (1996) 275.
%
%
\bibitem{paper4}  M.~L\"uscher, S.~Sint, R.~Sommer and H.~Wittig,
                  Nucl. Phys. B491 (1997) 344.
%
\bibitem{qmass} S. Capitani, M.~L\"uscher, R.~Sommer and H.~Wittig, DESY preprint,
                DESY-98-154, hep-lat/9810063.
%
\bibitem{roberto_pert} 
A.~Bucarelli, F.~Palombi, R.~Petronzio and A.~Shindler, 
hep--lat/9808005.
%
\end{thebibliography}
\end{document}